\documentclass[12pt]{article}
\input epsf.sty
\usepackage{epsfig}
\newcommand{\beq}{\begin{eqnarray}}
\newcommand{\eeq}{\end{eqnarray}}

\newcommand{\be}{\begin{equation}}
\newcommand{\ee}{\end{equation}}
\newcommand{\ben}{\begin{eqnarray}\displaystyle}
\newcommand{\een}{\end{eqnarray}}

\newcommand{\sectiono}[1]{\section{#1}\setcounter{equation}{0}}

\def\sqr#1#2{{\vcenter{\vbox{\hrule height.#2pt
         \hbox{\vrule width.#2pt height#1pt \kern#1pt
            \vrule width.#2pt}
         \hrule height.#2pt}}}}

\begin{document}

{}~ \hfill\vbox{\hbox{hep-ph/0309305} \hbox{PUPT-2097} }\break

\vskip 1cm

\begin{center}
\Large{\bf Is ${\bf \Theta^+ (1540)}$ a Kaon--Skyrmion Resonance?}

\vspace{20mm}

\normalsize{Nissan Itzhaki, Igor R. Klebanov, Peter Ouyang, and
Leonardo Rastelli}

\vspace{10mm}

\normalsize{\em Joseph Henry Laboratories, Princeton University,}

\vspace{0.2cm}

\normalsize{\em Princeton, New Jersey 08544, USA}
\end{center}

\vspace{10mm}

\begin{abstract}

\medskip

We reconsider the relationship between the bound state and the $SU(3)$
rigid rotator approaches to strangeness in the Skyrme model. For
non-exotic $S=-1$ baryons the bound state approach matches for small
$m_K$ onto the rigid rotator approach, and the bound state mode turns
into the rotator zero-mode. However, for small $m_K$, we find no
$S=+1$ kaon bound states or resonances in the spectrum, confirming
previous work. This suggests that, at least for large $N$ and small
$m_K$, the exotic state may be an artifact of the rigid rotator
approach to the Skyrme model. An $S=+1$ near-threshold state comes
into existence only for sufficiently large $SU(3)$ breaking. If such a
state exists, then it has the expected quantum numbers of $\Theta^+$:
$I=0$, $J=\frac{1}{2}$ and positive parity.  Other exotic states with
$(I=1, J^P=\frac{3}{2}^+)$, $(I=1,J^P=\frac{1}{2}^+)$,
$(I=2, J^P=\frac{5}{2}^+)$ and $(I=2,J^P=\frac{3}{2}^+)$
appear as its
$SU(2)$ rotator excitations. As a test of our methods, we also
identify a D-wave $S=-1$ near-threshold resonance that, upon $SU(2)$
collective coordinate quantization, reproduces the mass splittings of
the observed states $\Lambda(1520)$, $\Sigma(1670)$ and $\Sigma(1775)$
with good accuracy.

\end{abstract}

\newpage

\section{Introduction}

A remarkable recent event in hadronic physics is the discovery of a
$S=+1$ baryon (dubbed the $Z^+$ or $\Theta^+$) with a mass of
1540 MeV and width less than 25 MeV \cite{Nakano}. This discovery was
promptly
confirmed in \cite{Barmin,Stepanyan,Barth, Asratyan}.  At  present  the
spin, parity and magnetic moment of this state have not been
determined; one group, the SAPHIR collaboration \cite{Barth}, found
that the isospin of the $\Theta^+$ is zero.  Because it appears as a
resonance in the system $K^+ n$, the minimal possibility for its quark
content is $uudd\bar{s}$ which is manifestly exotic, {\it i.e.} it
cannot be
made out of three non-relativistic quarks. Early speculations on this
kind of exotic baryons were made in \cite{Jaffe,Strottman}.
Remarkably, a state with these quantum numbers appears naturally
\cite{Manohar,Chemtob,Praszalowicz} in the rigid rotator quantization
of the three-flavor Skyrme model \cite{Witten,Guad,MNP}, and detailed
predictions for its mass and width were made by Diakonov, Petrov and
Polyakov \cite{DPP}.  Their results provided motivation for the
experimental searches that led to the discovery of $\Theta^+$ very
close to the predicted parameters.\footnote{For other recent
theoretical models of $\Theta^+$, see,
e.g. \cite{Stancu,Karliner,JW,Cheung,Jennings,Csikor}.}

\smallskip

The rigid rotator quantization of the Skyrme model that was used in
\cite{DPP} relied on working directly with $N=3$ ($N$ is the number of
colors). Then the model predicts the well-known ${\bf 8}$ and ${\bf
10}$ of $SU(3)$, followed by an exotic ${\bf \overline {10}}$
multiplet whose $S=+1$ member is the $\Theta^+$.  The approach of
\cite{DPP} began by postulating that the established $N(1710)$ and
$\Sigma(1880)$ states are members of the anti-decuplet.  Then, using
group theory techniques, and constraints from a rigid rotator
treatment of chiral solitons, they estimated the mass and width of the
other states in this multiplet.  They predicted that the lowest member
of
the anti-decuplet has a mass of 1530 MeV and width of about 9
MeV.
These results appear to be confirmed strikingly by
experiment.\footnote{Note that the authors of
\cite{Nussinov,Arndt,Casher,Krein} have
argued that the experimental data actually indicate an even smaller
width.}

\smallskip

In this paper we will follow a somewhat different strategy, trying to
develop a systematic $1/N$ expansion for $\Theta^+$. In the two-flavor
Skyrme model,  the
quantum numbers of the low-lying states do not depend on $N$, as
long as it is odd: $I=J= \frac{1}{2}, \frac{3}{2}, \ldots$
($I$ is the isospin and and $J$ is the spin).
 These states are identified
with the nucleon and the $\Delta$ \cite{Skyrme, Adkins}.  The
three-flavor
case is rather different, since even the lowest $SU(3)$ multiplets
depend on $N=2n+1$ and become large as $N\to \infty$.  The allowed
multiplets must contain states of hypercharge $N/3$, {\it i.e.} of
strangeness $S=0$.  In the notation where $SU(3)$ multiplets are
labeled by $(p,q)$, the lowest multiplets one finds are $(1,n)$ with
$J=\frac{1}{2}$ and $(3,n-1)$ with $J=\frac{3}{2}$
\cite{Vadim,Dul,IRK,Cohen,DP}.
These are the large $N$ analogues of the octet and the decuplet. The
rigid rotator mass formula, valid in the limit of unbroken $SU(3)$, is
\begin{equation} \label{exactform}
M^{(p,q)}= M_{cl}+{1\over 2\Omega} J(J+1)
+{1\over 2\Phi} \left (C^{(p,q)} - J(J+1) -\frac{ N^2}{12} \right ) \, ,
\end{equation}
where $\Omega$ and $\Phi$ are moments of inertia, which are
of order $N$.
Using the formula for the quadratic Casimir,
\begin{equation}
C^{(p,q)} = {1\over 3} [ p^2+ q^2 + 3(p+q) + pq ]
 \, ,
\end{equation}
one notes \cite{Vadim,Dul,IRK,Cohen,DP}
that the lowest lying $SU(3)$ multiplets
$(2J, n+{1\over 2}- J)$, of spin $J={1\over 2}, {3\over 2}, \ldots$
obey the mass formula
\begin{equation} \label{lowest}
M(J)= M_{cl}+{N\over 4\Phi}+ {1\over 2\Omega} J(J+1)
\, .
\end{equation}
Exactly the same multiplets appear when we construct baryon states out
of $N$
quarks. The splittings among them are of order $1/N$,
as is usual for soliton rotation excitations.

\smallskip

The large $N$ analogue of the exotic antidecuplet is
$(0, n+2)$ with $J=\frac{1}{2}$. Its splitting from the lowest
multiplets is ${N\over 4\Phi} + O(1/N)$.
The fact that it is of order $N^0$ raises questions
about the validity of the rigid rotator approach to these
states \cite{KK,IRK,Cohen}. Indeed, we will argue that a better
approximation
to these states is provided by the bound state approach \cite{CK}.
In the bound state approach one departs from the rigid rotator
ansatz, and adopts more general kaon fluctuation profiles.
This has $O(N^0)$ effect on energies of states even in the
low-lying non-exotic multiplets, after $SU(3)$ breaking is turned on.
In the limit $m_K\to 0$ the bound state description of non-exotic
baryons smoothly approaches the rigid rotator description, and
the bound state wave function approaches the zero-mode
$\sin (F(r)/2)$ \cite{CK,CHK}, where $F(r)$ is the radial profile
function
of the skyrmion.

\smallskip

Our logic leads us to believe that, at least from the point of view of
the large $N$ expansion, $\Theta^+$ should be described by a
near-threshold
kaon-skyrmion resonance or bound state of $S=+1$, rather
than by a rotator state (a similar suggestion was
made independently in \cite{Cohen,Lebed}).
This is akin to the bound state
description of the $S=-1$ baryons in \cite{CK} where a possibility of
such a description of an exotic $S=+1$ state was mentioned as well.
This leads us to a puzzle, however, since, in contrast with the
situation for $S=-1$, for $S=+1$ there is no fluctuation mode that in
the $m_K\to 0$ limit approaches the rigid rotator mode of energy
${N\over 4\Phi}$ (this will be shown explicitly in section 3). An
essential difficulty is that, for small $m_K$, this is not a
near-threshold state; hence, it is not too surprising that it does not
show up in the more general fluctuation analysis.  Thus, for large $N$
and small $m_K$ the rigid rotator state with $S=+1$
appears to be an artifact of the rigid
rotator approximation (we believe this to be a general statement that
does
not depend on the details of the chiral lagrangian).

\smallskip

Next we ask what happens as we increase $m_K$ keeping other parameters
in the kaon-skyrmion Lagrangian fixed. For $m_K=495 MeV$, and with the
standard fit values of $f_\pi$ and $e$, neither bound states
\cite{CHK} nor resonances \cite{Scoc} exist for $S=+1$.\footnote{
 Such states seem to appear in \cite{KM}. However, in the
normalization of the Wess-Zumino term, which is repulsive for $S=+1$
states, a large factor of $e^2\sim 30$ was apparently omitted there
(compare eq. (7) of \cite{KM} with \cite{CHK}). When this factor is
reinstated, both bound states and resonances disappear for standard
parameter choices, as claimed in \cite{CHK,Scoc} and confirmed in this
work.}  If, however, we increase $m_K$ to $\approx 1 GeV$ then a
near-threshold state appears. Thus, we reach a surprising conclusion
that,
at least for large $N$, {\it the exotic $S=+1$ state exists only due
to the $SU(3)$ breaking and disappears when the breaking is too weak}.
While this certainly contradicts the philosophy of \cite{DP}, it is
actually in line with some of the earlier literature (see, for
example, \cite{Riska} and
the end of \cite{Weigel}). An intuitive way to see the
necessity of the $SU(3)$ breaking for the existence of the exotic is
that the breaking keeps it a near-threshold state.

\smallskip

One of the purposes of this paper is to examine how sensitive the
existence of this state is to parameter choices.  If we set
$m_K=495\,$MeV, then minor adjustments of $f_\pi$ and $e$ do not make
the $S=+1$ resonance appear.  We will find, however, that if the
strength of the Wess-Zumino term is reduced by roughly a factor of 0.4
compared to its $SU(3)$ value, then a near-threshold state corresponding
to $\Theta^+$ is indeed found.  Although we do not have a good a
priori explanation for this reduction, it could be caused by
unexpectedly large $SU(3)$ breaking effects on this particular
term. This issue clearly requires further investigation.

\smallskip

The structure of the paper is as follows. In the next section we will
review the rigid rotator approach to Skyrme model at large $N$, and
recall a method of large $N$ expansion introduced in \cite{KK}, which
involves expanding in rigid motions around the $SU(2)$ subgroup of
$SU(3)$.  In section 3 we proceed to the bound state approach that, by
introducing extra degrees of freedom, has $O(N^0)$ effect on baryon
spectra. In section 4 we review the status of $S=-1$ baryons
based on kaon-skyrmion bound states, and also study a near-threshold
D-wave resonance \cite{Scoc} that, upon
$SU(2)$ collective coordinate quantization,
reproduces the observed states $\Lambda(1520)$, $\Sigma(1670)$
and $\Sigma(1775)$ with good accuracy.
In section 5 we
carry out the search for $S=+1$ kaon-skyrmion bound states and
resonances.
In section 6 we attempt a different fit with a non-zero pion
mass. We offer some concluding remarks in section 7.

\sectiono{Three-flavor Skyrme model at large $N$}

The Skyrme approach to baryons begins with the Lagrangian \cite{Skyrme}
\beq
L_{Skyrme} = \frac{f^2_{\pi}}{16} {\rm Tr}(\partial_{\mu} U^{\dag}
\partial^{\mu} U)
+\frac{1}{32e^2}{\rm Tr}([\partial_{\mu} U U^{\dag},\partial_{\nu} U
U^{\dag}]^2)
+{\rm Tr}( M (U+U^{\dag}-2)) \, ,
\label{lskyrme}
\eeq
where $U(x^{\mu})$ is a matrix in $SU(3)$ and $M$ is proportional to
the matrix of quark masses.  Later on, it will be convenient to choose
units where $ef_{\pi}=1$.  There is an additional term in the action,
called the Wess-Zumino term:
\beq
S_{WZ} = -\frac{iN}{240\pi^2}\int d^5x
\epsilon^{\mu\nu\alpha\beta\gamma}
{\rm Tr}(\partial_{\mu} U U^{\dag}\partial_{\nu} U
U^{\dag}\partial_{\alpha} U U^{\dag}\partial_{\beta} U
U^{\dag}\partial_{\gamma} U U^{\dag})
\label{wz}.
\eeq
In the limit of unbroken $SU(3)$ flavor symmetry, its normalization
 is fixed by anomaly considerations \cite{WittenWZ}.

\smallskip

The Skyrme Lagrangian is a theory of mesons but it describes
baryons as well.  The simplest baryons in the Skyrme model are the
nucleons.  Classically, they have no strange quarks, so we may set the
kaon fluctuations to zero and consider only the $SU(2)$-isospin
subgroup of $SU(3)$.  Skyrme showed that there are topologically
stabilized static solutions of hedgehog form:
\beq
U_0 = U_{\pi,0} =\left(
\begin{array}{cc}
e^{i{ {\bf \tau} \cdot \hat{r}}F(r)} & 0\\
0 & 1 \end{array}\right)
\label{hedgehog}
\eeq
in which the radial profile function $F(r)$ satisfies the boundary
conditions $F(0)=\pi, F(\infty)=0$.  By substituting the hedgehog
ansatz (\ref{hedgehog}) into the Skyrme Lagrangian (\ref{lskyrme}),
and considering the corresponding equations of motion one obtains an
equation for $F(r)$ which is straightforward to solve numerically.
The non-strange low-lying excitations of this soliton are given by
rigid rotations of the pion field $A(t) \in SU(2)$:
\beq
U(x,t)=A(t)U_0 A^{-1}(t).
\label{rot}
\eeq
For such an ansatz the Wess-Zumino term does not contribute.
By expanding the Lagrangian about $U_0$ and canonically quantizing the
rotations, one finds that the Hamiltonian is
\beq
H=M_{cl} + \frac{1}{2\Omega} J(J+1) \, ,
\eeq
where $J$ is the spin and the c-numbers $M_{cl}$ and $\Omega$ are
complicated integrals of functions of the soliton profile.
Numerically, for vanishing pion mass, one finds that
\beq
M_{cl} &\simeq& 36.5 \frac{f_{\pi}}{e}, \label{mcl}\\
\Omega &\simeq& \frac{107}{e^3 f_{\pi}} \, .\label{omega}
\eeq
For $N=2n+1$, the low-lying quantum numbers are independent of the
integer $n$.  The lowest states, with $I=J=\frac{1}{2}$ and
$I=J=\frac{3}{2}$, are
identified with the nucleon and $\Delta$ particles respectively. Since
$f_\pi\sim \sqrt{N}$, and $e\sim 1/\sqrt{N}$, the soliton mass is
$\sim N$, while the rotational splittings are $\sim 1/N$.  Adkins,
Nappi and Witten \cite{Adkins} found that they could fit the $N$ and
$\Delta$ masses with the parameter values $e=5.45, f_{\pi}=129$ MeV.
In comparison, the physical value of $f_{\pi}=186$ MeV.

\smallskip

A generalization of this rigid rotator treatment that produces $SU(3)$
multiplets of baryons is obtained by making the collective coordinate
$A(t)$ an element of $SU(3)$.  Then the WZ term makes a crucial
constraint on allowed multiplets \cite{Witten,Guad,Manohar,Chemtob,MNP}.
As discussed in the introduction, large $N$ treatment of this 3-flavor
Skyrme model is
more subtle than for its 2-flavor counterpart. When $N=2n+1$ is large,
even the lowest lying $(1,n)$ $SU(3)$ multiplet contains $(n+1)(n+3)$
states with strangeness ranging from $S=0$ to $S=-n-1$
\cite{IRK}. When the strange quark mass is turned on, it will
introduce a splitting of order $N$ between the lowest and highest
strangeness baryons in the same multiplet.  Thus, $SU(3)$ is badly
broken in the large $N$ limit, no matter how small $m_s$ is
\cite{IRK}.  We will find it helpful to think in terms of $SU(2)\times
U(1)$ flavor quantum numbers, which do have a smooth large $N$ limit.
In other words, we focus on low strangeness members of these
multiplets, whose $I$, $J$ quantum numbers do have a smooth large $N$
limit, and to try  identify them with observable baryons.

\smallskip

Since the multiplets contain baryons with up to $\sim N$ strange
quarks, the wave functions of baryon with fixed strangeness deviate
only an amount $\sim 1/N$ into the strange directions of the
collective coordinate space. Thus, to describe them, one may expand
the $SU(3)$ rigid rotator treatment around the $SU(2)$ collective
coordinate.  The small deviations from $SU(2)$ may be assembled into a
complex $SU(2)$ doublet $K(t)$.  This method of $1/N$ expansion was
implemented in \cite{KK}, and reviewed in \cite{IRK}.

\smallskip

{}From the point of view of the Skyrme model the ability to expand in
small fluctuations is due to the Wess-Zumino term which acts as a large
magnetic field of order $N$. The method works for arbitrary kaon mass,
and has the correct limit as $m_K\to 0$.  To order $O(N^0)$ the
Lagrangian has the form \cite{KK}
\begin{equation}
L = 4\Phi \dot K^\dagger \dot K + i {N\over 2} (K^\dagger \dot K - \dot
K^\dagger  K )
- \Gamma K^\dagger K \, .
\end{equation}
The Hamiltonian may be diagonalized:
\begin{equation}
H= \omega_- a^\dagger a +
\omega_+ b^\dagger b + {N\over 4\Phi}
\ ,
\end{equation}
where
\begin{equation}
\omega_\pm = {N\over 8\Phi} \left (\sqrt{1+ (m_K/M_0)^2} \pm 1 \right )
\ ,\qquad M_0^2 = {N^2\over 16 \Phi\Gamma }\ .
\end{equation}
The strangeness operator is $S= b^\dagger b - a^\dagger a$.  All the
non-exotic multiplets contain $a^\dagger$ excitations only. In the
$SU(3)$ limit, $\omega_-\to 0$, but $\omega_+ \to {N\over 4\Phi}\sim
N^0$.  Thus, the ``exoticness'' quantum number mentioned in \cite{DP}
is simply $E= b^\dagger b$ here, and the splitting between multiplets
of different ``exoticness'' is ${N\over 4\Phi}$, in agreement with
results found from the exact rigid rotator mass formula
(\ref{exactform}).

\smallskip

The rigid rotator prediction for $O(N^0)$ splittings are not exact,
however, for reasons explained long ago \cite{CK,KK,IRK} and reviewed
in the introduction. Even for non-exotic states, as the soliton
rotates into strange directions, it experiences deformation which
grows with $m_K$.  The bound state approach allows it to deform, which
has a significant $O(N^0)$ effect on energy levels. We now turn
to review the bound state approach.

\sectiono{Review of the Bound State Approach}

Another approach to strange baryons, which proves to be quite
successful in describing the light hyperons, is the so-called bound
state method \cite{CK}.  The basic strategy involved is to expand the
action to second order in kaon fluctuations about the classical
hedgehog soliton.  Then one can obtain a linear differential equation
for the kaon field, incorporating the effect of the kaon mass, which
one can solve exactly.  The eigenenergies of the kaon field are then
precisely the differences between the Skyrmion mass and the strange
baryons.  In order to implement this strategy, it is convenient to
write $U$ in the form
\beq
U=\sqrt{U_{\pi}} U_K \sqrt{U_{\pi}} \, ,
\label{uform}
\eeq
where $U_{\pi}=\exp[2i\lambda_j \pi^j/f_{\pi}]$ and
$U_K=\exp[2i\lambda_a K^a/f_{\pi}]$ with $j$ running from 1 to 3 and
$a$ running from 4 to 7.\footnote{There is actually a second coupling
constant $f_K$ which replaces $f_\pi$ in the definition of $U_K$;
experimentally, $f_K\sim 1.22f_\pi$.  To incorporate the difference
between these coupling constants, one simply replaces $f_\pi$ by $f_K$
when expanding in powers of the kaon field, but does not rescale the
kinetic and kaon mass terms, which are required to have standard
normalization.  Then those terms that follow from the four-derivative
term, the Wess-Zumino term, and the pion mass in the Skyrme Lagrangian
change by a factor of $(f_\pi/f_K)^2$.}  The $\lambda_a$ are the
standard $SU(3)$ Gell-Mann matrices.  We will collect the $K^a$ into a
complex isodoublet $K$:
\beq
K=
\frac{1}{\sqrt{2}}\left(
\begin{array}{c}
K^4-iK^5 \\
K^6-iK^7
\end{array} \right)=
\left(
\begin{array}{c}
K^+ \\K^0
\end{array} \right).
\label{kdoublet}
\eeq
Though the Wess-Zumino term can only be written as an action term, if
we expand it to second order in $K$, we obtain an ordinary Lagrangian
term:
\beq
L_{WZ} = \frac{iN}{f_{\pi}^2}B^{\mu} \left( K^{\dag}
D_{\mu}K-(D_{\mu}K)
^{\dag}K \right)
\eeq
where
\beq
D_{\mu}K=
\partial_{\mu}K+\frac12\left(\sqrt{U_{\pi}^{\dag}}
\partial_{\mu}\sqrt{U_{\pi}}+\sqrt{U_{\pi}}
\partial_{\mu}\sqrt{U_{\pi}^{\dag}}\right)K \, ,
\eeq
and $B_\mu$ is the baryon number current. Now we decompose
the kaon field into a set of partial waves.  Because
the background soliton field is invariant under combined spatial and
isospin rotations ${\bf T} = {\bf I}+{\bf L}$, a good set of quantum
numbers is $T,L$ and $T_z$, and so we write the kaon eigenmodes as
\beq
K=k(r,t) Y_{TLT_z} \, .
\eeq
Substituting this expression into $L_{Skyrme}+L_{WZ}$ we obtain an
effective Lagrangian for the radial kaon field $k(r,t)$:
\beq \nonumber
L=4\pi \int r^2 dr\left( f(r)\dot{k}^{\dag}
\dot{k}+i\lambda(r)(k^{\dag}\dot{k}-\dot{k}^{\dag}k)-h(r)\frac{d}{dr}k^{\dag}
\frac{d}{dr}k -k^{\dag}k(m_K^2+V_{eff}(r)) \right) \, ,
\eeq
with $\lambda(r) = -\frac{N e^2}{2\pi^2 r^2}F'\sin^2 F$,
$f(r)=1+2s(r)+d(r),h(r)=1+2s(r), d(r)=F'^2, s(r)=(\sin F/r)^2,
c(r)=\sin^2\frac{F}{2}$, and
\beq
V_{eff}=-\frac14(d+2s)-2s(s+2d)+\frac{1+d+s}{r^2}(L(L+1)+2c^2+4c{\bf
I\cdot L})
\\+\frac{6}{r^2}\left(s(c^2+2c{\bf I\cdot L}-{\bf I\cdot
L})+\frac{d}{dr}\left((c+{\bf I\cdot L})F'\sin
F\right)\right)-\frac{m_{\pi}^2}{2}(1-\cos F) \, . \nonumber
\label{veff}
\eeq
The resulting equation of motion for $k$ is
\beq
-f(r)\ddot{k} +2i\lambda(r)\dot{k}+{\cal O} k=0\, , \\
{\cal O} \equiv \frac{1}{r^2}\partial_r
h(r)r^2\partial_r-m_K^2-V_{eff}(r)\, . \nonumber
\eeq
Expanding $k$ in terms of its eigenmodes gives
\beq
k(r,t)=\sum_{n>0}(\tilde{k}_n(r)e^{i\tilde{\omega}_nt}b_n^{\dag}+k_n(r)e^{i\omega_nt}a_n)\,
,
\eeq
with $\omega_n,\tilde{\omega}_n$ positive.  The eigenvalue equations are
thus
\beq
(f(r)\omega_n^2+2\lambda(r)\omega_n+{\cal O} )k_n&=&{ 0} \qquad(S=-1)\,
,\nonumber \\
(f(r)\tilde{\omega}_n^2-2\lambda(r)\tilde{\omega}_n+{\cal
O})\tilde{k}_n&=&
{ 0}
\qquad (S=+1)\, .
\label{keqns}
\eeq
Crucially, the sign in front of $\lambda$, which is the contribution
of the WZ term, depends on whether the relevant eigenmodes have
positive or negative strangeness.  The important result here is the
set of equations (\ref{keqns}) which we will now solve and whose
solutions we will match with the spectrum of baryons.

\smallskip

It is possible to examine these equations analytically for $m_K=0$.
The $S=-1$ equation has an exact solution with $\omega=0$ and
$k(r)\sim \sin (F(r)/2)$. This is how the rigid rotator zero mode is
recovered in the bound state treatment \cite{CHK}. As $m_K$
is turned on, this solution turns into an actual bound state
\cite{CK,CHK}. One the other
hand, the $S=+1$ equation does {\it not} have a solution with $\tilde
{\omega} ={N\over 4\Phi}$ and $k(r)\sim \sin (F(r)/2)$. This is why the
exotic rigid rotator state is not reproduced by the more precise bound
state approach to strangeness. In section 5 we further check
that, for small $m_K$, there is no resonance that would turn into
the rotator state of energy ${N\over 4\Phi}$ in the $SU(3)$ limit.

\sectiono{Baryons with $S=-1$}

In this section we recall the description of $S=-1$ baryons as
antikaon--skyrmion bound states or resonances.  We will set the kaon
mass equal to its physical value, $m_K=495$ MeV, but set the pion mass
equal to zero.  If we wish to fit both the nucleon and delta masses to
their physical values using the $SU(2)$ rotator approximation, then we
must take $e=5.45$ and $f_{\pi}=129$ MeV; let us begin with these
values as they are somewhat traditional in analyses based on the
Skyrme model.

\smallskip

The lightest strange excitation is in the channel $L$=1,
$T=\frac{1}{2}$, and
its mass is $M_{cl}+0.218\, ef_{\pi} \simeq 1019$ MeV.
As the lightest state
with $S=-1$, it is natural to identify it with the $\Lambda(1115)$,
$\Sigma(1190)$, and $\Sigma(1385)$ states, where the additional
splitting arises from $SU(2)$ rotator corrections.  Let us compute
these corrections.  The relevant formula for $S=-1$ \cite{CHK} is
\beq  \label{relevant}
M=M_{cl}+\omega_1+\frac{1}{2\Omega}\left[cJ(J+1)+(1-c)I(I+1)+\frac34
(c^2-c)\right]\, ,
\eeq
where $\omega_1$ is the kaon eigenenergy and $c$ is a number defined
in terms of the bound state eigenfunction $k_1$ by
\beq
c_{l=1}=1-\omega_1\frac{\int dr k_1^* k_1 \left(\frac43
fr^2\cos^2\frac{F}{2}-2(\frac{d}{dr}(r^2F'\sin F)-\frac43 \sin^2 F
\cos^2 \frac{F}{2}) \right)}{\int r^2 dr k_1^* k_1(f\omega_1+\lambda)}\,
.
\eeq
In this $L=1$, $T=\frac{1}{2}$ channel, we find from numerical
integration that
$c=0.617$.  The masses, including $SU(2)$ corrections, appear in
columns (a) of Table 1.  The two features to note here are that first,
these states are all somewhat
overbound, and second, the $SU(2)$ splittings match rather closely
with experiment (one of the successes of the bound state approach
\cite{CHK}).
\begin{table}
\begin{center}
\begin{tabular}{||r|r|r|r|r|r|r|r||} \hline
Particle& $J$& $I$& $L$& Mass (expt)& Mass (a)& Mass (b)& Mass (c)\\
\hline\hline
$\Lambda $& $\frac{1}{2}$ & 0& 1& 1115& 1048& 1059& 1121\\ \hline
$\Sigma   $& $\frac{1}{2}$& 1& 1& 1190& 1122& 1143& 1289\\ \hline
$\Sigma$& $\frac{3}{2}$& 1& 1& 1385& 1303& 1309& 1330\\ \hline
$\Lambda$& $\frac{1}{2}$& 0& 0& 1405& 1281& 1346& 1366\\ \hline
\end{tabular}
\end{center}
\caption{Masses (in MeV) of the light $S=-1$ hyperons as calculated from
the bound state approach, with (a) $e=5.45$, $f_\pi=f_K=129$ MeV, (b)
$e=4.82$, $f_\pi=f_K=186$ MeV, with an overall constant added to fit the
$N$ and $\Delta$ masses, and (c) the same parameters as (a) but with the
WZ term artificially decreased by a factor of 0.4.  In all cases
$m_\pi=0$.}
\end{table}

\smallskip

The next group of strange excitations is in the channel
$L=0$, $T=\frac{1}{2}$,
and we have determined its mass before including rotator corrections to
be
$M_{cl}+0.523 \, ef_{\pi}\simeq 1233$ MeV.
In this channel the formula for $c$ is given by
\beq
c_{l=0}=1-\omega_2\frac{\int dr k_2^* k_2 \left(\frac43
fr^2\sin^2\frac{F}{2}+2(\frac{d}{dr}(r^2F'\sin F)+\frac43 \sin^2 F
\sin^2 \frac{F}{2}) \right)}{\int r^2 dr k_2^* k_2(f\omega_2+\lambda)}
 \, ,
\eeq
and for the relevant bound state this gives $c\sim 0.806$.  The
$SU(2)$ corrections (\ref{relevant})
raise the mass of the lightest state in this
channel to 1281 MeV.  From the quantum numbers, it is natural to
identify this state with the $\Lambda(1405)$ state, but as we see it
is rather overbound.

\smallskip

We have seen that with the traditional values $e=5.45$ and
$f_{\pi}=129$ MeV the Skyrme model successfully captures qualitative
features of the baryon spectrum such as the presence of the
$\Lambda(1405)$ state, but that the bound states are all too light.
It is possible that the zero-point energy of
kaon fluctuations,  which is hard to calculate explicitly,
has to be added to all masses.  Thus it is easiest to
focus on mass splittings. Then from the $SU(2)$ rotator
quantization, we would obtain only one constraint (\ref{omega}), from
the nucleon-$\Delta$ splitting, and be able to adjust $e$ and
$f_{\pi}$ to improve the fit to known masses.  As we increase
$f_{\pi}$, we find that the particle masses increase, improving
agreement with experiment.  For definiteness, let us try to set
$f_{\pi}$ to its experimental value, $f_{\pi}$=186 MeV, which then
requires $e=4.82$.  We report the results for the masses in column (b)
of Table 1.

\smallskip

More dramatic increases in the particle masses may be obtained by
distinguishing between the pion decay constant $f_{\pi}$ and the kaon
decay constant $f_K$, as shown by Rho, Riska and Scoccola \cite{Rho},
who worked in a modified Skyrme model with explicit vector
mesons \cite{vector}.  For
$f_K=1.22\,f_{\pi}$ they were able to essentially eliminate the
over-binding problem for the $L=1$, $T=\frac{1}{2}$ states, though they
still
found the analogue of the $\Lambda(1405)$ state to be overbound by
about 100 MeV. We should add that the natural appearance of this
$\Lambda(1405)$ with negative parity is a major success of the
bound state approach \cite{CHK, Gobbi}.
In  quark or bag models such a baryon is described
by a $p$-wave quark excitation, which typically turns out
to be too heavy (for a discussion, see the introduction of
\cite{Gobbi}).

\begin{figure}[htbp]
\centering
\epsfig{file=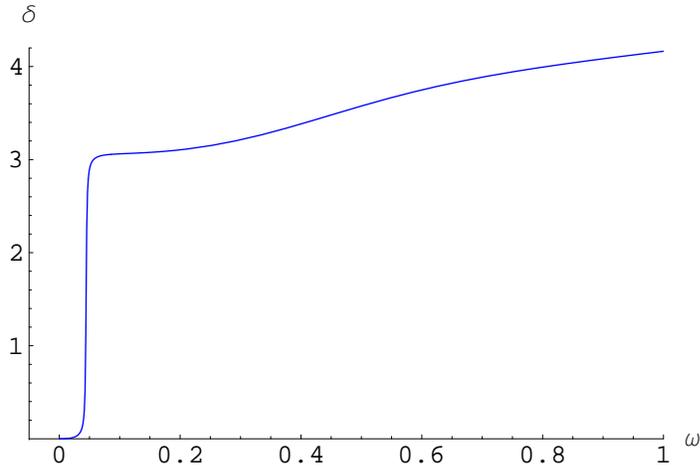, width=4in}
\caption{Phase shift as a function of energy in the
$L=2$, $T=\frac{3}{2}$, $S=-1$ channel.
The energy $\omega$ is measured in units of $ef_\pi$ (with the kaon mass
subtracted, so that $\omega=0$ at threshold), and
the phase shift $\delta$ is measured in radians.  Here $e=5.45$ and
$f_\pi=129$ MeV.}
\label{dwavefig}
\end{figure}

\smallskip

A third way to raise the masses of the $\Lambda$ and $\Sigma$
states is to modify the Wess-Zumino term.  For the $S=-1$ states,
the WZ term results in an attractive force between the Skyrmion
and the kaon, so if we reduce this term by hand, the hyperons will
become less tightly bound and their masses should
increase.\footnote{Note that
for unbroken $SU(3)$ the WZ term is quantized and
cannot be changed by hand. However, $SU(3)$ breaking is likely to
change the WZ term.} We will address this approach in the next
section, and we will see that such a reduction helps to produce an
$S=+1$ near-threshold state. However, too great a reduction of the WZ
term
will spoil the hyperfine splittings governed by the parameter $c$.

\smallskip
\begin{table}
\begin{center}
\begin{tabular}{||r|r|r|r|r|r||} \hline
Particle& $J$& $I$& $L$& Mass (expt) & Mass (th) \\ \hline\hline
$\Lambda \; \; (D_{03})$ & $\frac{3}{2}$ & 0& 2& 1520& 1462\\ \hline
$\Sigma \; \; (D_{13})$& $\frac{3}{2}$ & 1& 2& 1670& 1613\\ \hline
$\Sigma \; \; (D_{15})$& $\frac{5}{2}$ & 1& 2& 1775& 1723\\ \hline
\end{tabular}
\end{center}
\caption{Masses (in MeV) of the $S=-1$ $D$-wave resonances calculated
from the bound state approach, with $f_\pi = 129 $ MeV, $e = 5.45$. }
\end{table}

Finally, let us note that the philosophy of the bound state approach
can be applied successfully to states above threshold.  Such states
will appear as resonances in kaon-nucleon scattering, which we may
identify by the standard procedure of solving the appropriate kaon
wave equation and studying the phase shifts of the corresponding
solutions as a function of the kaon energy. In the $L=2$,
$T=\frac{3}{2}$ channel there is a resonance at $M_{cl}+0.7484 \,
ef_{\pi}$=1392
MeV (see Figure 1)\footnote{The existence of
this resonance was noted long ago in a
vector-meson stabilized Skyrme model~\cite{Scoc}.}.
In this channel, the $SU(2)$
splitting parameter $c$ is given by the formula \cite{Blom}
\beq
 \nonumber
c_{l=2}=1-\omega_3\frac{\int dr k_3^* k_3 \left(\frac23(1+\frac45
\cos^2\frac{F}{2})
fr^2-\frac{4}{5}(\frac{d}{dr}(r^2F'\sin F)+\frac43 \sin^2 F
\sin^2 \frac{F}{2}) \right)}{\int r^2 dr k_3^* k_3(f\omega_3+\lambda)}
\,.
\eeq
Numerically, we evaluate this coefficient by cutting off
the radial integral around the point where $k(r)$ begins to oscillate.
We find  $c \sim 0.23$.  The states are split into the channels with
$(I, J)$
given by  $(0, \frac{3}{2})$, $(1, \frac{3}{2})$,
$(1, \frac{5}{2})$ \cite{Scoc},
 with masses 1462 MeV, 1613
MeV, and 1723 MeV respectively (see Table 2).
We see that these correspond nicely to the known negative
parity resonances $\Lambda(1520)$ (which is $D_{03}$ in standard
notation),
$\Sigma(1670)$ (which is $D_{13}$) and
$\Sigma( 1775)$ (which is $D_{15}$) \cite{PDG}.
As with the bound states, we find that the resonances are
somewhat overbound (the overbinding of all states is presumably
related to the necessity of adding an overall zero-point energy
of kaon fluctuations), but
that the mass splittings within this multiplet are accurate to within a
few percent. In fact, we find that the ratio
\beq
{M(1, \frac{5}{2})- M(0, \frac{3}{2})\over
M(1, \frac{3}{2}) - M(0, \frac{3}{2})}
\approx 1.73
\ ,
\eeq
while its empirical value is $1.70$.

\smallskip

This very good agreement with the states observed above the $K-N$
threshold is an additional success of the kaon fluctuation approach to
strange baryons.

\section{A Baryon with S=+1?}

For states with positive strangeness, the eigenvalue equation for the
kaon field is the same except for a change of sign in the contribution
of the WZ term.  This sign change makes the WZ term repulsive for
states with $\bar{s}$ quarks and introduces a splitting between
ordinary and exotic baryons \cite {CK}.  In fact, with standard values
of the parameters (such as those in the previous section) the
repulsion is strong enough to remove {\em all} bound states and
resonances with $S=1$, including the newly-observed
$\Theta^+$.  It is natural to ask how much we must modify the Skyrme
model to accommodate the pentaquark.  The simplest modification we can
make is to introduce a coefficient $a$ multiplying the WZ term.
Qualitatively, we expect that reducing the WZ term will make the
$S=+1$ baryons more bound, while the opposite should happen to the
ordinary baryons.  Another modification we will attempt is to vary the
mass of the kaon; we will find that for sufficiently large kaon mass
the $\Theta^+$ becomes stable.  In all cases, we have found
empirically that raising $f_K$ relative to $f_\pi$ makes the
pentaquark less bound, so for this section we will take $f_K=f_\pi$.

\smallskip

The most likely channel in which we might find an exotic has the
quantum numbers $L=1$, $T=\frac{1}{2}$,
as in this case the effective
potential is least repulsive near the origin.  For $f_{\pi}$= 129,
186, and 225 MeV, with $e^3f_{\pi}$ fixed, we have studied the effect
of lowering the WZ term by hand.  Interestingly, in all three cases we
have to set $a\simeq 0.39$ to have a bound state at threshold.  If we
raise $a$ slightly, this bound state moves above the threshold, but does
not
survive far above threshold; it ceases to be a sharp state for $a\simeq
0.46$.  We have plotted phase shifts for various values of $a$ in Figure
2.\footnote{When the state is above the threshold, we do not
find a full $\pi$ variation of the phase. Furthermore,
the variation and slope of the phase shift
decrease rapidly as the state moves higher, so it
gets too broad to be identifiable. So, the state can only exist as a
bound
state or a near-threshold state.
}
With $a=0.39$ and $f_\pi=129$ MeV, this state (with mass essentially at
threshold) has $SU(2)$ splitting parameter $c\sim -0.48$. The $SU(2)$
collective coordinate
quantization of the state proceeds analogously to that of the $S=-1$
bound states, and the mass formula is again of the form
(\ref{relevant}).
  Thus the lightest $S=+1$ state we find has $I = 0$, $J = \frac{1}{2}$
and positive parity, {\it i.e.} it is an $S= +1$ counterpart of the
$\Lambda$.
This is our candidate $\Theta^+$ state.
Its first $SU(2)$ rotator excitations have $I =1$, $J^P = \frac{3}{2}^+$
and $I =1$, $J^P = \frac{1}{2}^+$ (a relation of these
states to $\Theta^+$ also follows from
general large $N$ relations among baryons \cite{Lebed}).
The counterparts of these
$J^P=\frac{1}{2}^+, \frac{3}{2}^+$
states in the rigid rotator quantization lie in the {\bf 27}-plets of
$SU(3)$.
These states were recently discussed in
\cite{Walliser, Kobushkin}.

From the mass formula (\ref{relevant}) we deduce that
\begin{equation}
M(1, \frac{1}{2})- M(0, \frac{1}{2}) ={1\over \Omega}(1-c)
\ ,\qquad
M(1, \frac{3}{2})- M(0, \frac{1}{2}) ={1\over \Omega}
\left (1+{c\over 2}\right )\ .
\end{equation}
Since $c < 0$, the $J = \frac{3}{2}$ state
is lighter than $J = \frac{1}{2}$.
 Using $c\sim -0.48$, we find
that the $I=1$, $J^P = \frac{3}{2}^+$ state is $\sim 148$ MeV
heavier than the $\Theta^+$,\footnote{This value is close
to those predicted in \cite{Walliser} but is significantly higher
than the $55$ MeV reported in \cite{Kobushkin}. For a comparison of
these possibilities with available data, see
\cite{Jennings}.}
while the $I =1$, $J^P = \frac{1}{2}^+$
state is $\sim 289$ MeV
heavier than the $\Theta^+$.
%Perhaps these states may be detectable
%experimentally via their decays into a pion and a $\Theta^+$.

We may further consider $I=2$ rotator excitations which have
$J^P= \frac{3}{2}^+, \frac{5}{2}^+$.
Such states are allowed for $N=3$ (in the quark
language the charge $+3$ state, for example, is given by
$uuuu{\bar s}$).
The counterparts of these
$J^P=\frac{3}{2}^+, \frac{5}{2}^+$
states in the rigid rotator quantization lie in the {\bf 35}-plets of
$SU(3)$ \cite{Walliser, Kobushkin}.
From the mass formula (\ref{relevant}) we deduce that
\beq
& M(2, \frac{5}{2})- M(0, \frac{1}{2}) ={1\over \Omega}(3+c)\sim 494
\ {\rm MeV}\ ,\\
& M(2, \frac{3}{2})- M(0, \frac{1}{2}) ={3\over \Omega}
\left (1-{c\over 2}\right )\sim 729\ {\rm MeV}\ . \nonumber
\eeq

While the value of $c$ certainly depends on
the details of the chiral lagrangian, we may form certain combinations
of masses of the exotics from which it cancels. In this way we find
``model-independent relations'' which rely only the existence of
the $SU(2)$ collective coordinate:
\beq \label{modelin}
2 M(1, \frac{3}{2}) + M(1, \frac{1}{2})- 3 M(0, \frac{1}{2})
& =&
2(M_\Delta- M_N)=586\ {\rm MeV} \ \nonumber ,\\
\frac{3}{2} M(2, \frac{5}{2}) + M(2, \frac{3}{2})- \frac{5}{2}
M(0, \frac{1}{2}) & =& 5(M_\Delta- M_N)= 1465\ {\rm MeV}\ ,  \\
     M(2,\frac{3}{2}) - M(2,\frac{5}{2}) & =& \frac{5}{3} \left(
M(1,\frac{1}{2})-  M(1,\frac{3}{2}) \right) , 
\nonumber 
\eeq
where we used $M_\Delta-M_N=\frac{3}{2\Omega}$. These relations
are analogous to the sum rule \cite{CK} \beq 2 M_{\Sigma^*} +
M_\Sigma- 3 M_\Lambda= 2(M_\Delta- M_N) \ , \eeq which is obeyed
with good accuracy. If the $I=1,2$ exotic baryons are discovered,
it will be very interesting to compare the relations
(\ref{modelin}) with experiment.
\smallskip

\begin{figure}[htbp]
\centering
\epsfig{file=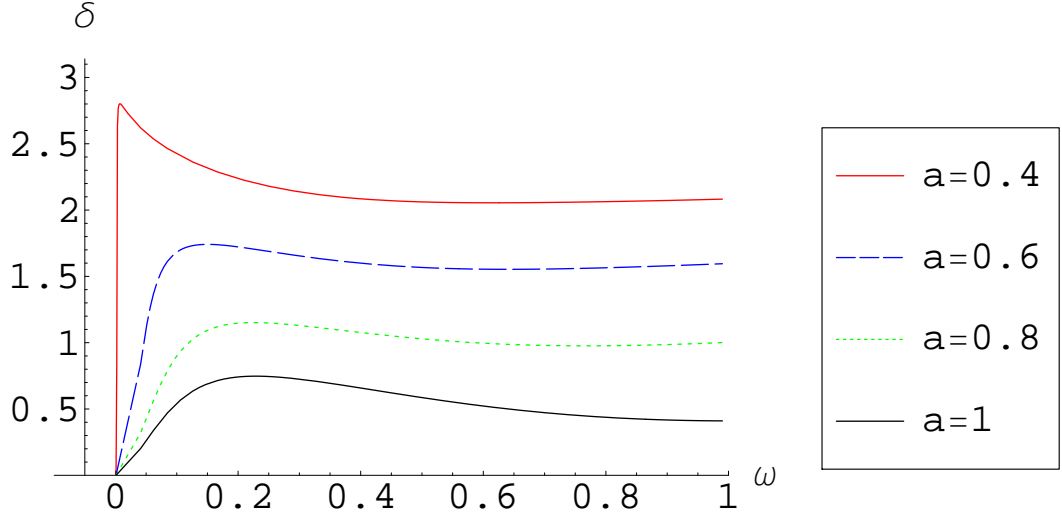, width=6in}
\caption{Phase shifts $\delta$ as a function of energy
in the $S=+1$, $L=1$, $T=\frac{1}{2}$ channel, for
various choices of the parameter $a$ (strength of the WZ term).
 The energy $\omega$ is measured in units of $ef_\pi$ ($e=5.45$,
$f_\pi=f_K= 129$ MeV) and
the phase shift $\delta$ is measured in radians. $\omega =0$
corresponds to the $K-N$ threshold.}
\label{wzplot}
\end{figure}

\begin{figure}[htbp]
\centering
\epsfig{file=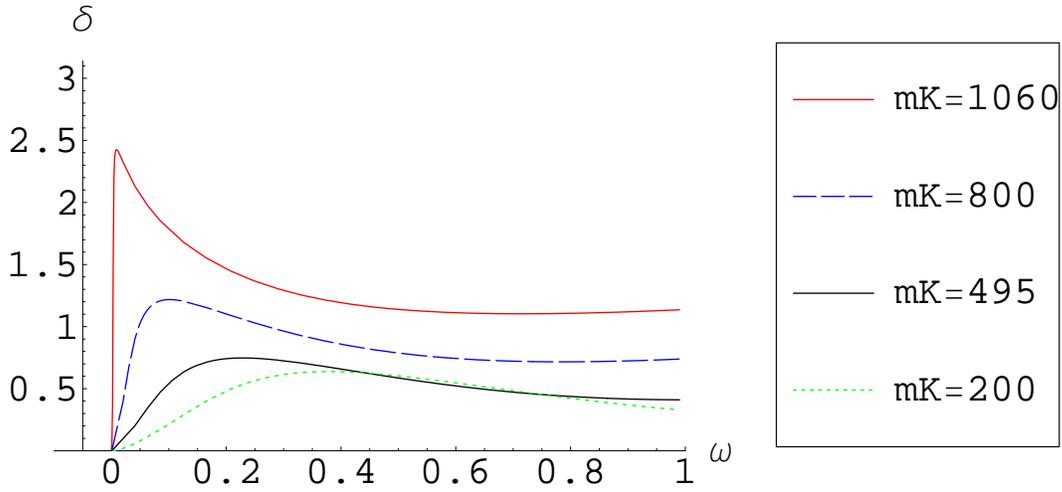, width=6in}
\caption{Phase shifts $\delta$ as a function of energy
in the $S=+1$, $L=1$, $T=\frac{1}{2}$ channel,
for various values of $m_K$.
Here $e=5.45$ and $f_\pi=f_K=129$ MeV.}
\label{mkplot}
\end{figure}
Suppose we tried to require the existence of a pentaquark state by
setting $a$=0.4.  How would this change affect the spectrum of
negative strangeness baryons?  The results are somewhat mixed.  Let us
take $f_{\pi}$=129 MeV as an example.  In the $L=1,T=\frac{1}{2}$
channel,
the
bound state has a mass of 1209 MeV before including $SU(2)$ rotator
corrections.  The parameter $c$ which characterizes the $SU(2)$
splittings falls to $c \sim 0.14$.  Including the splittings we find
the masses given in column (c) of Table 1.  Notice that the $\Sigma$
is far above its experimental mass of 1190 MeV, signaling drastic
disagreement with the Gell-Mann-Okubo relations, as we would expect
for this small value of $c$.  In the $L=0,T=\frac{1}{2}$ channel the
$\Lambda$ resonance is at 1366 MeV (including rotator corrections),
still somewhat overbound.

\smallskip

As another probe of the parameter space of our Skyrme model, let us
vary the mass of the kaon and see how this affects the pentaquark.  As
observed in Section 3, in the limit of infinitesimal kaon mass, there
is no resonance in the $S=+1, L=1, T=\frac{1}{2}$ channel.  We find
that to obtain a bound state in this channel, we must raise $m_K$ to
about 1100 MeV.\footnote{Since both $D$ and $B$ mesons are much
heavier than this, we may infer following \cite{Riska} that there
exist exotic bound state baryons which, in the quark model language,
are pentaquarks containing an anti-charm or an anti-bottom quark
(provided that the associated meson decay constants $f_D$ and $f_B$
are not too large). This prediction is rather insensitive to the
details of the chiral lagrangian \cite{Riska}. See also \cite{oh}
for a more careful analysis of these exotics that incorporates
heavy quark symmetry.}  Plots of the phase
shift vs. energy for different values of $m_K$ are presented in Figure
3. This figure clearly shows that increasing $SU(3)$ breaking leads to
increasing variation of the scattering phase.
%Notice howevever
%that one obtains a true resonance (phase shift $\sim \pi$)
%for values of $m_K$

\section{Fits with Massive Pion}

One interesting way to extend the model is to include the mass of the
pion, as first explored by Adkins and Nappi \cite{an}.
We find that this actually improves matters for the pentaquark.
The basic procedure is simply to take $m_{\pi}=138$ MeV; there will be
a modification in the kaon effective potential and also a change in
the variational equation for the Skyrmion profile function $F(r)$.  As
a result, the constraints on $e$ and $f_{\pi}$ change:
\beq
M_{cl} &\simeq& 38.7 \frac{f_{\pi}}{e}, \label{mclpi}\\
\Omega &\simeq& \frac{62.9}{e^3 f_{\pi}}.\label{omegapi}
\eeq
Adkins and Nappi found that the best fit to the nucleon and delta masses
was given by $e=4.84$ and $f_{\pi}=108$ MeV.  In dimensionless units,
$m_{\pi}=.263$.

\smallskip

For the $S=-1$ baryons, we find a bound state in the $L=1$,
$T=\frac{1}{2}$ channel with mass 1012 MeV.  The $SU(2)$ rotator
parameter is $c_{l=1}\sim 0.51$, giving a mass spectrum of 1031, 1126,
and 1276 MeV.  In the $L=0$, $T=\frac{1}{2}$ sector there is a bound
state with mass 1204 MeV which is increased by rotator corrections
($c\sim 0.82$) to give a state with mass 1253 MeV corresponding
to $\Lambda(1405)$.
Furthermore, in the $L=2$, $T=\frac{3}{2}$ channel
there is a bound state slightly below the threshold
\cite{Blom} (in the massless
pion fit this state was a resonance slightly above the
threshold). While these
states are all overbound, one can presumably improve the fit by
adjusting parameters as in the previous sections.

\smallskip

Let us first note that with the massive pion fit, the pentaquark
appears with a smaller adjustment in parameters.  In the $L=1$,
$T=\frac{1}{2}$,
$S=+1$ channel, there is a bound state for $a \sim 0.69$. Let us study
the effect of setting $a=0.75$ on the bound states.  With this change,
the bound state energies for negative strangeness baryons rise.  In
the $L=1$, $T=\frac{1}{2}$
channel, the mass is 1114 MeV, and $c_{l=1}\sim
0.35$, giving masses of 1050, 1177 and 1279 MeV for the
$\Lambda,\Sigma,\Sigma$ states.  $c_{l=1}$ is smaller than its
experimental value of 0.62 but is not disastrously small, and the
overall masses have increased towards their experimental values.  In
the $L=0, T=\frac{1}{2}$ channel,
the $\Lambda$ state also increases to 1283
MeV (including rotator corrections.)
\begin{table}
\begin{center}
\begin{tabular}{||r|r|r|r|r|r|r||} \hline
Particle& $J$& $I$& $L$& Mass (expt)& Mass (d)& Mass (e)\\ \hline\hline
$\Lambda $& $\frac{1}{2}$ & 0& 1& 1115& 1031& 1050\\ \hline
$\Sigma  $& $\frac{1}{2}$& 1& 1& 1190& 1126& 1177\\ \hline
$\Sigma$& $\frac{3}{2}$ & 1& 1& 1385& 1276& 1279\\ \hline
$\Lambda$& $\frac{1}{2}$ & 0& 0& 1405& 1253& 1283\\ \hline
\end{tabular}
\end{center}
\caption{Masses (in MeV) of the light $S=-1$ hyperons as calculated from
the bound state approach, with $e=4.84$, $f_\pi=f_K=108$ MeV, and
$m_\pi=138$ MeV.  Column (d) reports the masses with the usual WZ term;
in column (d) the WZ term has been artificially reduced by a factor
0.75.}
\end{table}
One further interesting result is that the pentaquark seems more
sensitive to the mass of the kaon with $m_\pi=138$ MeV; the pentaquark
actually becomes bound for $m_K=700$ MeV.

\smallskip

It is also possible to vary the pion mass; for masses around 200 MeV the
bound state can appear for $S$=1 with $a\sim 0.8$.

\section{Discussion}

There are several implications of the analysis carried
out in the preceding sections. First of all, we have to admit that the
bound state approach to the Skyrme model could not have been used to
predict
the existence of an exotic $S=+1$ baryon. Indeed, for typical parameter
choices we find neither kaon-Skyrmion bound states nor resonances with
$S=+1$,
confirming earlier results from the 80's \cite{CHK,Scoc,IRK}.
At the time these results appeared consistent with the apparent absence
of
such exotic resonances in kaon scattering data.\footnote{In fact, the
experimental situation is still somewhat confused (see
\cite{Nussinov,Arndt,Casher,Krein} for discussions of remaining
puzzles).}

\smallskip

We have found, however, that by a relatively large adjustment of
parameters in the minimal bound state lagrangian, such as reduction of
the WZ term to $0.4$ of its $SU(3)$ value, the near-threshold $S=+1$
kaon state can be made to appear. In this case, however, the agreement
of the model with the conventional strange baryons is worsened
somewhat. A better strategy may be to vary more parameters in the
Lagrangian, and perhaps to include other terms; then there is hope
that properties of both exotic and conventional baryons will be
reproduced nicely.  This would be a good project for the future.

\smallskip

Finally, our work sheds new light on connections between the bound
state and the rigid rotator approaches to strange baryons. These
connections were explored in the 80's, and it was shown that the bound
state approach matches nicely to 3-flavor rigid rotator quantization
carried out for large $N$ \cite{CHK,KK,IRK}. The key observation is
that, in both approaches, the deviations into strange directions
become small in the large $N$ limit due to the WZ term acting as a
large magnetic field. Thus, for baryons whose strangeness is of order
$1$, the harmonic approximation is good for any value of $m_K$. The
$S=-1$ bound state mode smoothly turns into the rotator zero-mode in
the limit $m_K\to 0$, which shows explicitly that rotator modes can be
found in the small fluctuation analysis around the $SU(2)$ skyrmion.
However, for small $m_K$ there is no fluctuation mode corresponding to
exotic $S=+1$ rigid rotator excitations. In our opinion, this confirms
the seriousness of questions raised about such rotator states for
large $N$ \cite{KK,IRK,Cohen}.

\smallskip
If the large $N$ expansion is valid, then we conclude that the exotic
baryon appears in the spectrum only for sufficiently large $SU(3)$
breaking. The simplest way to parametrize this breaking is to increase
$m_K$ while keeping coefficients of all other terms fixed at their
$SU(3)$ values. Then we find that the resonance appears at a value of
$m_K \sim 1 GeV$. However in reality $SU(3)$ breaking will also affect
other coefficients, in particular it may reduce somewhat the strength
of the WZ term, thus helping the formation of the resonance. In
principle the coefficients in the chiral lagrangian should be fitted
from experiment, and also higher derivative terms may need to be
included.

\smallskip

What does our work imply about the status of the $\Theta^+$ baryon in
the real world? As usual, this is the most difficult question.  If
$N=3$ is large enough for the semiclassical approach to skyrmions to
be valid, then we believe that our picture of $\Theta^+$ as a
kaon-skyrmion near-threshold state is a good one. It is possible,
however, that the rigid rotator approach carried out directly for
$N=3$, as in \cite{DPP}, is a better approximation to the real world,
as suggested by its successful prediction of the pentaquark.  It is
also possible that quark model approaches, such as those in
\cite{Karliner,JW}, or lattice calculations \cite{Csikor}, will
eventually prove to be more successful.  Clearly, further work, both
experimental and theoretical, is needed to resolve these issues.

%%%%%%%%%%%%%%%%%%%%%%%%%%%%%%%%%%%%%%%%%%%%%%%
\section*{Acknowledgments}
%%%%%%%%%%%%%%%%%%%%%%%%%%%%%%%%%%%%%%%%%%%%%%%%%%%%
We are grateful to C. Callan, D. Diakonov,
G. Farrar, C.Nappi, V. Petrov and E. Witten for useful
discussions.
This material is based upon work
supported by the National Science Foundation Grants No.
PHY-0243680 and PHY-0140311.
Any opinions, findings, and conclusions or recommendations expressed in
this material are those of the authors and do not necessarily reflect
the views of the National Science Foundation.

\begingroup\raggedright\endgroup
\end{document}